# Discussion on the origin of magic numbers in clusters

Siwei Luo[1]

[1]Department of Chemical and Biomolecular Engineering, Ohio State University, Columbus, OH 43210

**Abstract:** The distribution of the sizes of clusters is not continuous, but rather has local maxima. The numbers of atoms of those maxima distribution is called magic numbers. Two methods of determining magic numbers are firstly introduced, followed by three different models which were developed to explain the origin of magic numbers. Close-packing better explain those clusters build up with regular shells; LJ potential was used to calculate the energy properties of clusters which partially meet with the occurrence of magic numbers; LJ-plus-AT or LJ-plus-EX give a more detailed analysis of interaction mechanism.

## 0 Introduction:

The concept of "magic numbers" in physics was not born in connection with clusters of atoms or molecules, the subject of this contribution. It originated in the late forties, when Maria Goeppert-Mayer discovered on an empirical basis that nuclei with certain numbers of protons or neutrons(2,8,20,50,82, and 126; the last number refers to neutrons only) are particularly stable(Fig 1). The liquid-drop model and the uniform model proved to be inherently incapable of explaining such discontinuities. The same author approached this remarkable phenomenon herself theoretically by assuming that strong spin-orbit forces exist, giving rise to a sequence of independent particle states which match the experimentally observed irregularities. Practically simultaneously, Haxel, Jensen, and Suss observed the same phenomena and developed the same interpretation. The expression "magic numbers" came into use almost instantly. These analyses, and those by several other nuclear physicists during the same period, led to the development of the nuclear shell model. Although the adjective "magic" suggests phenomena "seemingly requiring more than human power; startling in performance; producing effects which seem supernatural"(Webster's New College Dictionary), this non-scientific label has been adopted through-out as serving the purpose of identifying these numbers simply and conveniently.

Let us turn to atomic or molecular clusters. For non-metallic clusters, size distributions have been measured using nozzle-beam techniques, or electron diffraction. An intriguing observation has been that often such distributions are not continuous, but are marked by local maxima. Usually, such a maximum appears gradually with increasing cluster size, after which a steep decrease takes place. Fig 2 a, b, c show the magic numbers exist in different clusters, such as lithium, Na, Ar, and protonated water clusters and so forth. Following the nuclear physicists of the forties, it has become customary to call these numbers n* of maximal intensity "magic numbers".

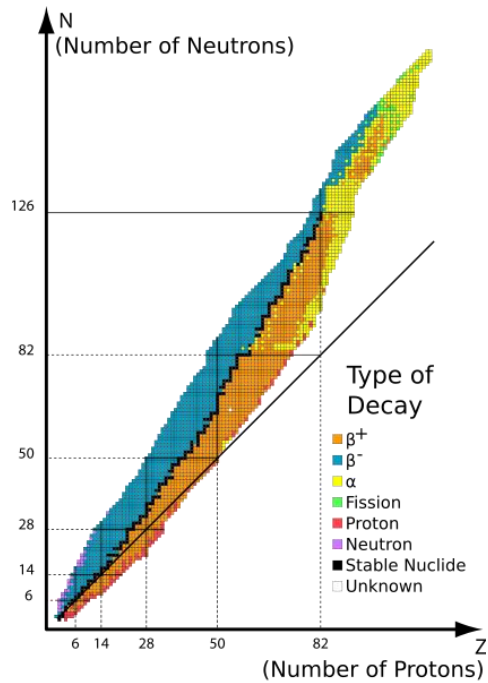

**Fig 1** Number of neutrons versus number of protons. Adapted from: http://en.wikipedia.org/wiki/Stable_nuclide

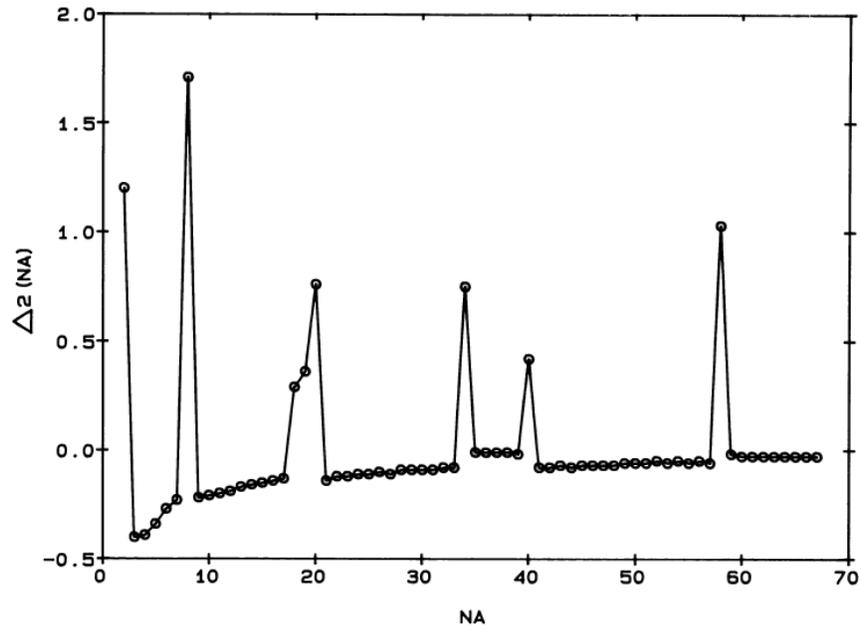

**Fig 2a** Second energy difference [$\Delta_2$(NA)] for lithium clusters versus number of atoms (NA).[1-4]

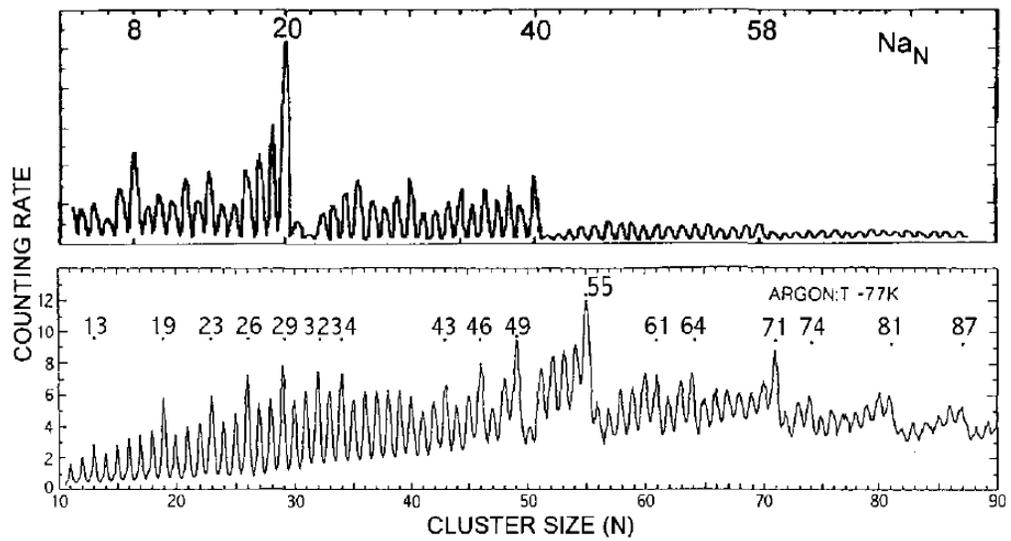

**Fig 2b** Mass spectra measured for Ar and Na clusters. The intense peaks indicate enhanced stability.[1-4]

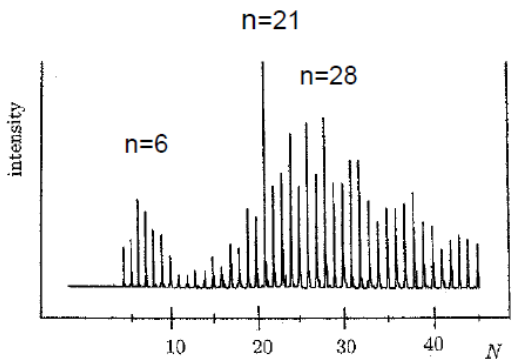
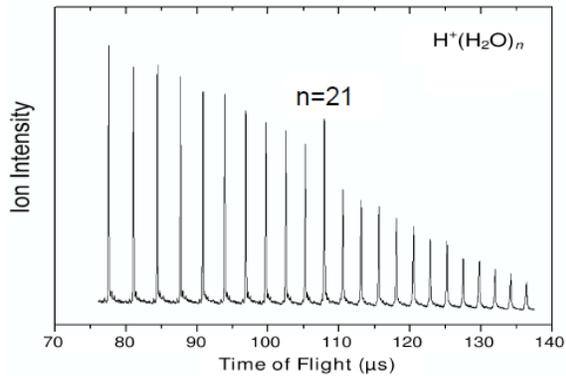

Fig 2c magic number at n=6,21, and possibly at 24, 26, 28 in the mass spectra of $H^+(H_2O)_n$ clusters[1-4]

# 1 Method of measurement

Theoretically, the accurate evaluation of the concentration Cn of an n-particle cluster constitutes one of the principal, and most difficult problems in nucleation physics. Especially in nozzle-beam applications the experimental situation is so complicated that it is practically impossible to retrace the thermal history of a given observed cluster. However, two different avenues of approach can be sketched:

A) At the time of measurement, the ensemble of clusters may, in good approximation, be treated as a multi-component system in thermodynamic equilibrium. The concentration Cn of a cluster of n particles in then given by Boltzmann distribution

Cn = C1 exp[-ΔG(n,p,T)/kBT]

Where C1 is independent of n, at given pressure p and temperature T. The quantity ΔG is the free enthalpy (Gibbs free energy) of formation of the cluster from n isolated particles. Plotting ΔG(n,p,T) against n, maxima in Cn correspond to "dips" in the free enthalpy of formation.

B) At the time of measurement, the ensemble of clusters in neither in thermodynamic equilibrium nor in a steady state, i.e. Cn depends on the time t. The time evolution of the size distribution is, in the absence of condensation, and assuming monomer evaporation to be dominant,

dCn/dt=Rn+1Cn+1- RnCn

where Rn is the evaporation rate of a cluster of n particles. The quantity Rn is simply taken proportional to $\exp[-(\Delta G(n-1)- \Delta G(n))/k_BT]$, where we have omitted the variables p, T attached to ΔG. The difference $\Delta G(n-1)- \Delta G(n)$ will be called "sublimation free enthalpy" of the cluster of n monomers. Maxima in the cluster-size distribution will now correspond to maxima in the sublimation free enthalpy as a function of cluster size n, at given p and T.

The general theoretical problem of finding maxima in the cluster-size distribution (i.e. the magic numbers ) turns to find "either type", i.e. either those associated with dips in the free enthalpy of formation ΔG(n,p,T) or those corresponding to maxima in the sublimation free enthalpy.

## 2  Different models

To get rid of confusing complications, the system we considered is restricted to thermally stabilized, neutral, rare-gas clusters. To further simplify the problem, the clusters are assumed to be near zero K and zero-point energies are neglected. Under those assumptions, the potential energy determines relative stability of clusters and their most favorable configurations. And based on those assumptions, several different models have been developed historically.

## 2.1 Close packings of rigid spheres model

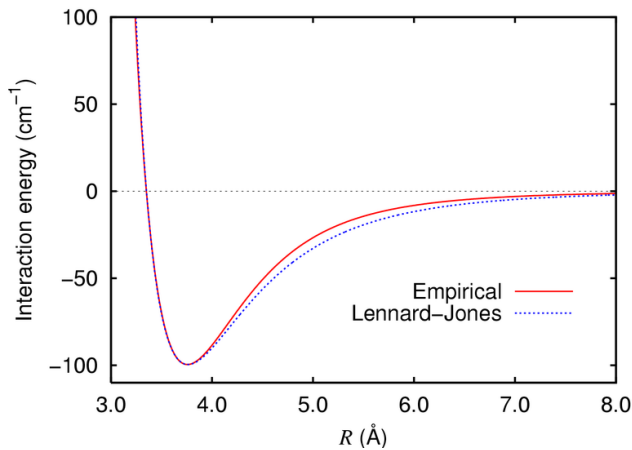

$$V(r) = 4\epsilon \left[\left(\frac{\sigma}{r}\right)^{12} - \left(\frac{\sigma}{r}\right)^{6}\right]$$

Fig 3 Lennard-Jones potential[1-4]

For rare-gas atoms, their interactions can be accurately demonstrated by a Lennard-Jones potential $4\varepsilon[(\sigma/R)^{12}-(\sigma/R)^{6}]$, where ε is the depth of the potential well, σ is the (finite) distance at which the inter-particle potential is zero, and r is the distance between the particles. These parameters can be fitted to reproduce experimental data or accurate quantum chemistry calculations. The $r^{-12}$ term describes Pauli repulsion at short ranges due to overlapping electron orbitals and the $r^{-6}$ term describes attraction at long ranges (van der Waals force, or dispersion force).

In nozzle-beam experiments the clusters are in a region of very low temperatures. Then the probability for two atoms to find themselves in the repulsive part of their potential is very small. It is thus tempting to replace this part by an infinitely steep wall, i.e. the atoms are rigid spheres interacting through an attractive $r^{-6}$ potential. That's the model of close packings of rigid spheres.

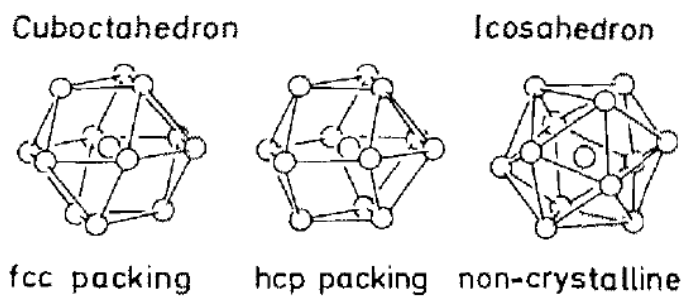

Fig 4 Face-centered cubic (fcc), hexagonal close packing (hcp) and the n=13 icosahedron[1-4]

The start magic number for both argon and xenon, which is n*=13 can be readily explained by the above three close packing styles. In fcc and hcp, the central atom is surrounded by 12 nearest neighbors, and an icosahedron, being more spherical, can better serve as explanation.

When more shells of the packing are added to predict other magic numbers, the results are not so satisfactory. The series are 13, 55, 147, 309, 561, …….Although the first three are indeed magic numbers for xenon, the numbers in between cannot be explained. Farge et al, proposed to fill these gaps by developing interpenetrating icosahedral structure. For instance, n*=19 is interpreted in terms of a double icosahedron sharing seven atoms. However, as to n*=25, this assumption fails.

Close packing of rigid spheres, plus $r^{-6}$ attraction, does offer some insight into the observed magic numbers, but this insight is much too crude to be considered sufficient.

## 2.2 Considering realistic pair potentials

In replacing the rigid-sphere by a Lenneard-Jones $r^{-12}$ repulsion, we hope to obtain more "structure" in the stability series. And since there is a negative outcome of analysis based on the traditional oscillator / rigid rotor (HO/RR) approximation, a combination of classical and quantum Monte Carlo methods is introduced. The argon atoms interact through a Lennard-Jones potential; internal energies, free energies, and entropies were calculated as a function of pressure and of cluster size. The results are plotted as follows:

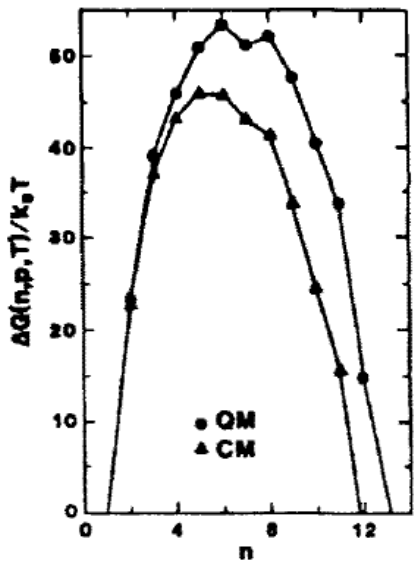
Fig 5a

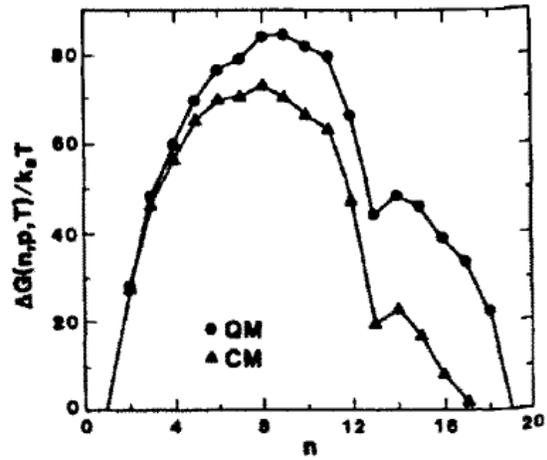
Fig 5b

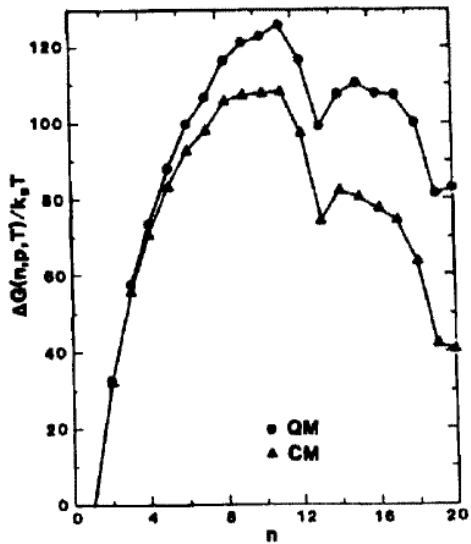
Fig 5c

Figure 5a, 5b, 5c. The Gibbs energy of formation as a function of argon cluster size at T=10K, and p=33.4fatm(5a), 0.334fatm(3b), and 3.34atm(3c), respectively. The circles are the quantum results and the triangles are the classical results.[1-4]

It can be concluded that magic numbers associated, at pressure p and fixed temperature T, with a dip in the free enthalpy of formation, emerge or disappear with pressure in a rather erratic way. Besides, the dips which do occur correspond to maxima in the sublimation energy of the clusters. However, the theoretically predicted magic number n*=7, corresponding to a pentagonal bipyramid on a Lennard-Jones basis, has not been found experimentally.

## 2.3 LJ-plus-AT & LJ-plus-EX

Lennard-Jones potential has provided a relatively solid basis for explaining the occurrence of maxima in cluster densities and of micro-cluster static morphology at low temperature. It is not surprising that practically all micro cluster calculations involving rare-gas atoms are based on this type of interaction.

However, one fundamental flaw with the Lennard-Jones potential does exist, which is: the stability of the face-centered cubic bulk-crystal structure cannot be explained on this basis. To solve this problem, short-range, many-atom interactions ( at least three-atom) have to be considered.. A simplified model of these is triple-dipole interaction (AT potential); and later on, three-atom exchange potential theory was proposed. Here is a brief discussion of the results calculated by these models.

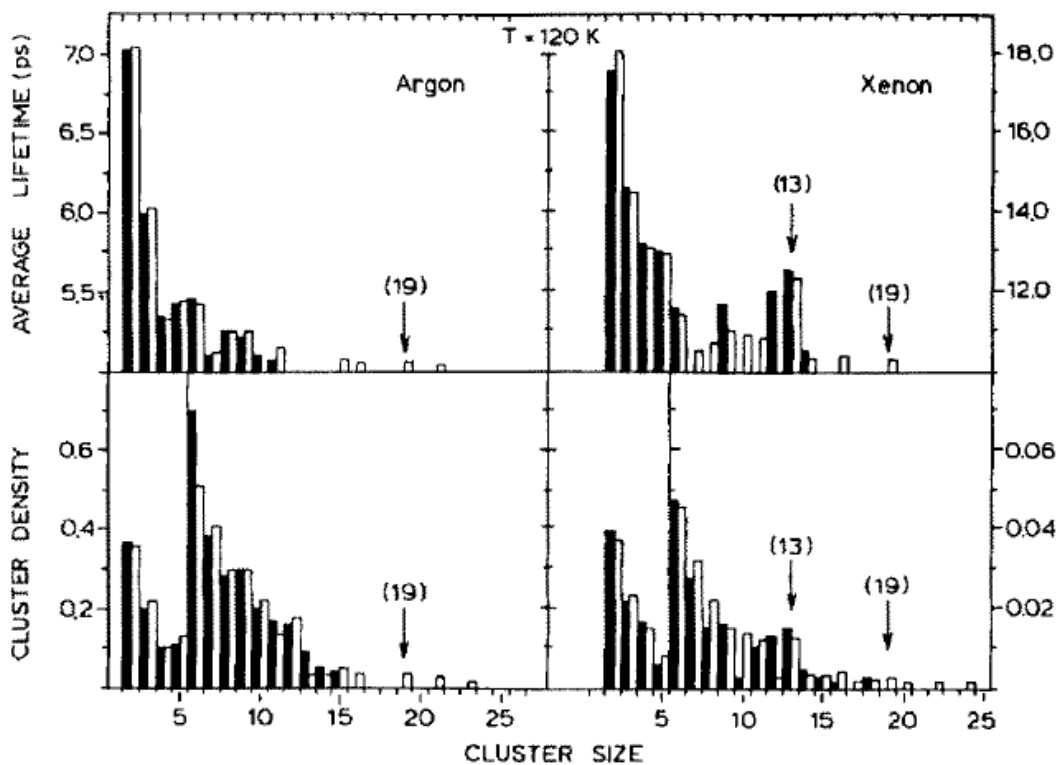

Fig 6 cluster density versus cluster size distribution (bottom) and average lifetimes (top) at a temperature of T=120K and reduced densities of 0.0553 and 0.0088 for Ar and Xe, respectively. [1-4]

The general observation from the figure is that the LJ-plus-EX potential leads to more stable clusters than LJ-plus-AT potential, even though the temperature is high. This is, in particular, reflected in the relative densities of clusters corresponding to "magic numbers" compared to the densities of their immediate neighbors.

## 3 conclusions

Based on the two methods of measurement of the occurrence of magic numbers, several models of cluster building have been introduced, including close-packing of solid sphere, Lennard-Jones potential, LJ-plus-AT, LJ-plus-EX models. Close-packing better explain those clusters build up with regular shells; LJ potential was used to calculate the energy properties of clusters which partially meet with the occurrence of magic numbers; LJ-plus-AT or LJ-plus-EX give a more detailed analysis of interaction mechanism. However, none of the theories are thought to be perfectly illustrating the mystery of magic numbers. Morphology, forms of internal forces, and representation of measurement, complexity of algorithm should all been taken into consideration when developing more "fine structured" theory of the origin of magic numbers. It is noteworthy to see that the magic cluster model theory has been utilized for many applications. For example, in Au nanoparticle catalysis, the growth of gold was modelled with a magic cluster model such that a certain growth pattern is only stable for a certain sized nanoparticle with a layer by layer growth fashion. This serves as the quantification standard for precise calculation such as surface palladium atoms deposition, thus allowing for accurate control of catalysis and kinetics.[5-21]